\documentclass[sigconf,nonacm]{acmart}
\AtBeginDocument{%
  }

\usepackage{listings}
\usepackage{courier}
\usepackage{longtable}
\usepackage{verbatim}
\usepackage{graphicx}
\usepackage{subcaption}

\begin{document}

\title{Trouble at the top: can Python extend the chains of trust in infrastructure firmware?}
\titlenote{Presented in the Works-in-Progress track of the 2026 Hot Topics in the
  Science of Security Symposium (HotSoS 2026), April 14--16, 2026.
  Session details: \url{https://sos-vo.org/group/hotsos/2026/hernandez};
  full agenda: \url{https://sos-vo.org/group/hotsos/agenda}. This is the
  authors' own version of the work, posted for non-commercial scholarly
  dissemination.}

\author{Larry Hernandez}
\orcid{1234-5678-9012}
\affiliation{%
  \institution{Dartmouth College}
  \city{Hanover}
  \state{New Hampshire}
  \country{USA}}

\author{Sergey Bratus}
\affiliation{%
  \institution{Dartmouth College}
  \city{Hanover}
  \state{New Hampshire}
  \country{USA}}

\renewcommand{\shortauthors}{Hernandez et al.}

\begin{abstract}

Compiled Python bytecode (PYC) has become an essential part of 
network switches, routers, and other network infrastructure devices.
Our analysis shows that its integrity is implicitly trusted in multiple designs
that make use of Python code at the top of the operational software, such as the
management and control pane of enterprise network switches.
At the same time, the integrity of PYC files is not covered under the
traditional chain-of-trust models, due to complex interactions with the
CPython loader, byte compiler, and other Python runtime components. We
explore the risks inherent in including PYC and Python runtimes in the
de facto trusted code basis of commercial enterprise equipment and offer a comprehensive
framework for understanding emergent behaviors in these designs.

\end{abstract}

\keywords{Python, Bytecode, PYC, PEP 552, Import System, Persistence, Software Supply Chain}

\maketitle

\section{Introduction}

Python has evolved from a scripting language that readably and clearly glued together
parts of projects and provided syntactically convenient access to the heavier C/C++-coded
components of the compute infrastructure, into a language in which major features of the
infrastructure itself (such as routing, switching, and network management) are now implemented. Python is convenient enough, maintainable
enough, and composable enough to save the infrastructure programmers
valuable time---and fast enough for most features outside of the hot paths. While Python
is not unique in these properties, it also commands a huge supply of available talent.
Economically speaking, its use in infrastructure is inescapable.

This shift in Python's role has a huge architectural implication for trust: Python doesn't
just glue together parts of a typical Trusted Code Base, it is a part of that TCB, however
trust is defined.

We must therefore consider Python's place in the system trust and integrity models such
the Chain of Trust, Trusted Path Execution (TPE), and Mandatory Access Control (cf. FLASK~\cite{smalleyflask}, SELinux, AppArmor, CHERI).
In particular, we must consider {\em integrity} of Python's executable artifacts and the behavior of
its {\em loaders} and runtime.  Moreover, since Python is transparently byte-compiled, the
behavior of {\em bytecode caches} and their lookup logic must all become a part of the
trust model analysis.  Since a chain of trust is a chain of loaders, parsers/deserializers, and
resolvers (even as trivial as looking up the next portion of code to measure and load in a
table or via a symbol label), all of these behaviors must be studied from the trust
engineering perspective \cite{saydjari2018engineering}.\footnote{Notably, Go, while having similar programmer economies to
Python, takes a very different approach, with large static compiled images \cite{cesarano2025goleash}.
}

Figuratively, Python is already found at the top of the chain of trust; can
it trustworthily link up with it?

Architecturally, the problem is larger that Python: how can empirically effective notions
of integrity and trust be extended to a dynamic environment with its own complex
translation, modularity, loading, and caching behaviors? \cite{loscocco1998inevitability}

We believe that the following is a first-class research problem: {\em can an interpreter support TPE policies
and extend the chain of trust?} With this paper, we hope
to open it up for discussion.

\subsubsection*{Paper structure}
We start with a survey of Python's existing mechanisms of module loading and byte compilation (Section~\ref{s:how-now}), attacks on these mechanisms (Section~\ref{s:attacks})
and why they matter for the computing infrastructure (Section~\ref{s:why-this-matters}), including a case
study that reveals patterns of vulnerabilities and persistence threats~(Sections~\ref{s:case-study}, \ref{s:persistence}). We
then discuss systems architecture implications (Section~\ref{s:architecture}) and future work (Section~\ref{s:future}).

\subsubsection*{\bf Our contributions}

In this work, we present a systematic analysis of Python bytecode loading across
major CPython versions, with particular emphasis on the structure and semantics
of \texttt{.pyc} headers and their role in import-time trust decisions. These
headers are the metadata that would drive any TPE-like policy for the Python
interpreter. 

We explicitly enumerate the dependencies of the timestamp-based and hash-based
invalidation modes of Python byte-compiled code (PYC) in the filesystem and on the OS path name resolution.

We demonstrate that the coexistence of timestamp-based and hash-based
invalidation modes, combined with permissive loader behavior, creates a
{\em practical post-compromise persistence} surface, which is a strong
concern for wherever Python is used in infrastructure of trusted integrity.

We show that
hash-based protections can be bypassed either by downgrading headers to
timestamp-based formats or by flipping a single flag bit to force unchecked-hash
mode, causing CPython to treat modified bytecode as authoritative without source
verification. While an explicit \textit{always} flag exists for the interpreter,
we also review how its use is virtually non-existent.

We then discuss the potential for extending traditional trust models to Python and point
out potential pitfalls involving mutable storage, package managers, user virtual
environments and other broadly used ecosystem features that all affect behaviors of the
loader, often differently at different program points where loading happens.

\section{Python modules, bytecode, and integrity evolution}
\label{s:how-now}

CPython's module import system \cite{cannon_import_internals,cpython_import}
relies heavily on precompiled bytecode files
(\texttt{.pyc}) for performance and deployment flexibility. The CPython interpreter
is capable of compiling Python source into standalone bytecode files. These are
generated during installation for the standard and built-in libraries, and whenever
a module is imported or used by another program.

As compiled bytecode is essentially cached (via the filesystem and multiple pieces of loader
logic), the {\em cache invalidation} problem immediately looms.
While the evolution of Python bytecode invalidation mechanisms---from timestamp-based validation to
hash-based schemes introduced in PEP~552\cite{pep552}---was motivated by determinism and
correctness, these mechanisms were not designed as actual security or trust chain boundaries. Yet since these mechanisms are implicitly involved in trust decisions, they need
to be analyzed from the trust perspective.

\subsection{Evolution of PYC headers}

In order to understand the impact and interactions of bytecode loading in CPython
with operating system subsystems and application-specific properties (such as mutable storage,
package managers, user virtual environments, etc.), it is crucial to review how the bytecode
invalidation mechanisms evolved over time.

These changes are primarily reflected in the source code and the successive PEP
(Python Enhancement Proposals) documents \cite{pep552}\cite{pep3147}\cite{pep3149}. The most substantial changes were introduced
between the 2.x and 3.7 major versions, culminating with the adoption of PEP~552 \cite{pep552} and
the implementation of hash-based invalidation.

The actual Python VM bytecode \cite{bendersky_python_internals,python_language_reference}
follows the header as a marshalled PyCodeObject \cite{cpython_marshal}. CPython itself does not guarantee a stable
consistent bytecode reference \cite{python_design_faq} across different versions.

\subsubsection{Python 2.x, 3.0--3.2}

The very early PYC header, up to 3.2 version of the CPython interpreter, notably
excluded any information beyond a timestamp (source file modification time). ``Timestomping'', i.e.,
malicious tampering with the timestamp to mimic that of the source file, sufficed to force
the malicious planted compiled bytecode file as authoritative over the source file.

\begin{table}[h]
\centering
\begin{tabular}{ccc l}
\toprule
Offset & Size & Field & Description \\
\midrule
0 & 4 & magic\_number & Version-specific magic \\
4 & 4 & timestamp & Source mtime \\
8 & N & code object & Marshalled PyCodeObject \\
\bottomrule
\end{tabular}
\caption{PYC header format in Python 2.x--3.2}
\end{table}

\subsubsection{Python 3.3--3.6}

From CPython 3.3 onwards, and until 3.6, an additional field (source file size in bytes)
was added, with no impact nor benefit in terms of security. Thus, the same trivial ``timestomping'' path for
forcing the PYC file as authoritative existed, solely based upon timestamp forgery on the
PYC file itself.

\begin{table}[h]
\centering
\begin{tabular}{ccc l}
\toprule
Offset & Size & Field & Description \\
\midrule
0 & 4 & magic\_number & CPython magic \\
4 & 4 & timestamp & Source mtime \\
8 & 4 & source\_size & Size of \texttt{.py} \\
12 & N & code object & Marshalled PyCodeObject \\
\bottomrule
\end{tabular}
\caption{PYC header format in Python 3.3--3.6}
\end{table}

\subsubsection{Python 3.7--3.12 Multi-mode Invalidation}

With the release of CPython 3.7%
the successive changes brought upon by the 
implementation of PEP~552, support for overlapping header structures was introduced.
This provided multi-mode support, effectively allowing compatibility with both the
legacy timestamp-based invalidation scheme (Table~\ref{tab:timestamp-based-pyc-312}),
and the new hash-based scheme (Table~\ref{tab:hash-based-pyc-312}).

The choice is mediated at runtime through the flags field. Because the data type size
of the 32-bit timestamp and source size fields is equal to that of the hash (64-bit),
no relocation of the marshalled PyCodeObject is needed.

\begin{table}[h]
\centering
\begin{tabular}{ccc l}
\toprule
Offset & Size & Field & Description \\
\midrule
0 & 4 & magic\_number & CPython magic \\
4 & 4 & flags & 0 = timestamp-based \\
8 & 4 & timestamp & Source mtime \\
12 & 4 & source\_size & Source size \\
16 & N & code object & Marshalled PyCodeObject \\
\bottomrule
\end{tabular}
\caption{Timestamp-based PYC (Python 3.7--3.12)}
\label{tab:timestamp-based-pyc-312}
\end{table}

\begin{table}[h]
\centering
\begin{tabular}{ccc l}
\toprule
Offset & Size & Field & Description \\
\midrule
0 & 4 & magic\_number & CPython magic \\
4 & 4 & flags & Bit 0: hash; Bit 1: checked \\
8 & 8 & hash\_value & SipHash of source \\
16 & N & code object & Marshalled PyCodeObject \\
\bottomrule
\end{tabular}
\caption{Hash-based PYC format (PEP~552)}
\label{tab:hash-based-pyc-312}
\end{table}

The header structure altogether is expressed as pseudocode in
Listing~\ref{lst:cpython-header-multimode-header}.

\begin{lstlisting}[label={lst:cpython-header-multimode-header},
  caption={CPython Multi-mode Invalidation header}]
uleshort    magic_number
string      "\x0d\x0a"
ulelong     flags
union {
  struct {
    ulelong timestamp
    ulelong size
  }
  ulequad   hash
}
\end{lstlisting}

The hash value is calculated via SipHash, as described in PEP~552:

\begin{quotation}
  We will use a SipHash with a hardcoded key of the contents of the source file.
  ...
  We choose SipHash because Python already has a builtin implementation of it from PEP 456,
  although an interface that allows picking the SipHash key must be exposed to Python.
  Security of the hash is not a concern, though we pass over completely-broken hashes
  like MD5 to ease auditing of Python in controlled environments.
  ...
\end{quotation}

The \path{_imp_source_hash_impl} function implements the actual keyed hashing internally,
in the CPython \path{import.c} source file \cite{cpython_import}. The key itself is not secret,
nor there is any protection for its value. It is known and defined within the CPython interpreter
at compile time (\path{include/internal/pycore_magic_number.h}).

\subsection{CPython PYC header handling}

Listing~\ref{lst:cpython-header-assemble-pyc} contains the reduced functional code
responsible for building the PYC header for both timestamp and hash based invalidation
schemes.

\begin{lstlisting}[language=Python, label={lst:cpython-header-assemble-pyc},
  caption={CPython helper functions for generating PYC files}]
def _code_to_timestamp_pyc(code, mtime=0,
  source_size=0):
    data = bytearray(MAGIC_NUMBER)
    data.extend(_pack_uint32(0))
    data.extend(_pack_uint32(mtime))
    data.extend(_pack_uint32(source_size))
    data.extend(marshal.dumps(code))
    return data

def _code_to_hash_pyc(code, source_hash,
  checked=True):
    data = bytearray(MAGIC_NUMBER)
    flags = 0b1 | checked << 1
    data.extend(_pack_uint32(flags))
    data.extend(source_hash)
    data.extend(marshal.dumps(code))
    return data
\end{lstlisting}

The header is immediately followed by the marshaled---serialized---Python bytecode
stream.

Quoting verbatim from PEP~552 \cite{pep552}:

\begin{quotation}
Runtime configuration of hash-based pyc invalidation is facilitated by the
\path{--check-hash-based-pycs} interpreter option. This is a tristate option,
which may take three values: \path{default}, \path{always}, and \path{never}.
\textbf{The default value, \path{default}, means the \texttt{check\_source} flag in
hash-based pycs determines invalidation as described above}. The
\path{always} setting causes the interpreter to hash the source file for
invalidation regardless of the value of the \texttt{check\_source} bit. The
\path{never} setting causes the interpreter to always assume hash-based pycs
are valid.

When \path{--check-hash-based-pycs=never} is in effect, unchecked hash-based
pycs will be \textbf{regenerated as unchecked hash-based pycs}. Timestamp-based pycs
are unaffected by \path{--check-hash-based-pycs}.
\end{quotation}

Emphasis is our own. This complexity in loader semantics is further complicated
by PEP~3147 \cite{pep3147}, which introduced the capability for multiple PYCs to be
co-located with the Python source file (.py file), and via PEP~552, in its absence.

Cursory analysis of the extent of the use of the \path{always} state for 
\path{--check-hash-based-pycs} reveals that it has had no adoption whatsoever
in publicly observable codebases. This was verified through a Github search excluding
copies of the PEP document, which amounted to the near totality of the references
to the option.

\section{Practical attacks via PYC invalidation}
\label{s:attacks}

It is important to note the context and applicability of the attacks described below.
We assume post-compromise access sufficient to modify bytecode files and filesystem metadata.
This reflects realistic attacker capabilities in user-space compromises and supply-chain
infection scenarios. Moreover, persistence is receiving increasing attention in actual cyber operations.

These conditions are almost invariably met in all the observed scenarios of relevance:
control or managenement pane applications in embedded systems (such as enterprise networking
equipment) typically run with superuser--root---privileges, user-controlled virtual environments are
mutable--writable---by the user himself, and CI (continuous integration) and build systems have full
access to the objects being built. Therefore, and especially in the context of post-compromise
persistence, the ability to tamper with filesystem objects and their metadata is typically already
present.

\subsection{Invalidation method downgrade (reversion to timestamp)}

Because of the similarities in the PYC headers across invalidation methods, it is possible
to force a regression from hash-based invalidation to the legacy timestamp-based scheme.

Provided that an adversary can
\begin{itemize}
    \item already alter the target PYC file, and
    \item issue system calls to alter file object properties (timestamps),
\end{itemize}

a seamless downgrade attack on the PYC header can be performed to effectively convert it
into a timestamp-based equivalent. This negates the benefits of the hash-based verification.
In case the original source file is missing, this will have an equivalent effect to the
"unchecked hash" technique described next.

\subsection{Hash invalidation bypass through unchecked-hash bit flip}
\label{sec:unchecked-hash-bypass}

PEP~552 defines an unchecked-hash mode in which CPython does not recompute or
verify the source hash. Flipping a single flag bit is sufficient to force this
mode, causing CPython to treat the bytecode as authoritative even when the source
is missing or inconsistent.

As established by the PEP 552 changes, the following conditions are present:

Flag bit pattern in the header:

\begin{itemize}
\item flags \& 0x01 == 1 $\longrightarrow$ hash-based
\item hash\_flags \& 0x01 == 0 $\longrightarrow$  unchecked
\end{itemize}

Once the unchecked mode flag is enabled:

\begin{itemize}
\item CPython does not recompute the source hash on import.
\item CPython does not even check for a .py file at all unless explicitly asked.
\end{itemize}

Therefore, \textbf{the .pyc is treated as authoritative}. The immediate effect is that the .pyc loads
even if the source code has changed, is missing, or is inconsistent with the embedded hash, without
alerting the user or raising any observable exception.

The legitimate expected use-cases are all unsurprisingly tangential to adversarial scenarios:

\begin{itemize}
\item Shipping “bytecode-only” applications
\item Environments where the source is stripped or unavailable
\item Embedded systems and constrained deployments
\item Performance-sensitive loaders (avoid hash recomputation)
\end{itemize}

These are especially aligned with the constraints of embedded systems, requiring both raw
performance and short boot times (time required to reach to operational state).

\section{Novel attack contributions}
\label{s:novel-attacks}

We provide a complete exploit workflow, tooling support, and a real-world case
study involving Python-based control planes in enterprise network equipment.
Our results illustrate that Python bytecode invalidation remains a correctness
feature rather than a security mechanism, and that defensive assumptions to the
contrary are unsafe.

\subsection{Manipulation toolkit: pycmangle}

We implement these techniques in \texttt{pycmangle}, a utility capable of parsing,
dumping, downgrading, and manipulating PYC headers. The tool focuses exclusively
on loader semantics rather than bytecode rewriting~\cite{lyneRemappingPythonOpcodes2020}, or infection of the marshaled
bytecode stream~\cite{patrascuWhenPythonsBite2016,barJgeralnikPytroj2025,BytecodeBytecode0171dev9+g13032bbdc}.

The toolkit implements the following functionality for PYC files:

\begin{itemize}
    \item Parsing and identification of the generating CPython interpreter, dumping the headers as-is.
    \item Manipulation of timestamp-based headers to "timestomp" the PYC file.
    \item Downgrading of hash-based headers for the same purpose (re-basing the bytecode stream), into timestamp-based headers.
    \item Downgrading of hash-based headers via unchecked mode flag bit swap.
\end{itemize}

All the techniques implemented serve the same purpose: making a tampered (modified)
PYC file authoritative in the eyes of the interpreter, overriding any existent `.py`
source file.

\subsection{Case study: Enterprise network equipment}
\label{s:case-study}
This research originated from an investigation into Python-based management
planes in Ruckus (formerly Brocade) FastIron ICX switches. Persistent Python bytecode stored
in flash partitions enables long-lived implants that survive reboots and factory
resets across multiple hardware generations.

Ruckus (formerly Brocade and FastIron) ICX switches employ a pseudo-package system
within their proprietary firmware image format. Packages sometimes contain Python
applications that are installed to a persistent flash partition. Infection or
tampering of the bytecode files could then be used to achieve de facto persistence,
even across factory resets, across some 15 years worth of enterprise switch models.

This was the subject of a presentation by the author at DistrictCon 2025 in Washington DC.

\subsection{FLAPPYSWITCH vulnerabilities in context}

FLAPPYSWITCH \cite{SubreptionFLAPPYSWITCH2025} demonstrated how multiple design and implementation flaws in the embedded OS FastIron
trust and update infrastructure can be chained together to achieve persistent code execution
and compromise of critical networking gear, even those evaluated against FIPS and Common
Criteria standards.

FastIron OS, a custom Linux-based operating system used across many ICX series routers and
switches, carries longstanding architectural weaknesses in its firmware verification
and trust chain.

The FLAPPYSWITCH research identified and leveraged several critical vulnerabilities in
FastIron’s firmware integrity and update infrastructure. Key issues include:

\begin{itemize}
\item CVE-2024-50606 \cite{hernandezSubreptionRuckusNetworks2025}: A fundamentally flawed trust chain in the FastIron boot and firmware
verification process, where integrity checks occur after code has already been loaded into
memory and rely on insecure TOCTOU (time-of-check, time-of-use) patterns.

\item CVE-2024-50607 \cite{hernandezCVE202450607RuckusNetworks2025}: Path traversal flaws permitting manipulation of command parameters.

\item CVE-2024-50604 \cite{hernandezCVE202450604RuckusNetworks2025}: Broken UFI package integrity verification, also suffering from TOCTOU
weaknesses.

\item CVE-2024-50605 \cite{hernandezCVE202450605RuckusNetworks2025}: A seemingly low-impact bug that enables arbitrary file writes,
but which provides a vector for filesystem corruption and persistence.
\end{itemize}

Collectively, these vulnerabilities allow an attacker with (at minimum) CLI breakout
or similarly limited access to escalate to full system compromise and persistence,
undermining FIPS/CC assurances. The monolithic architecture
of FastIron---with no effective privilege separation---amplified this issue, making almost
any subsystem a potential escalation vector.

\section{Why this matters: Analyzing the prevalence of Python programs in infrastructure firmware}
\label{s:why-this-matters}

\begin{figure}[t]
  \centering
    \includegraphics[width=\linewidth]{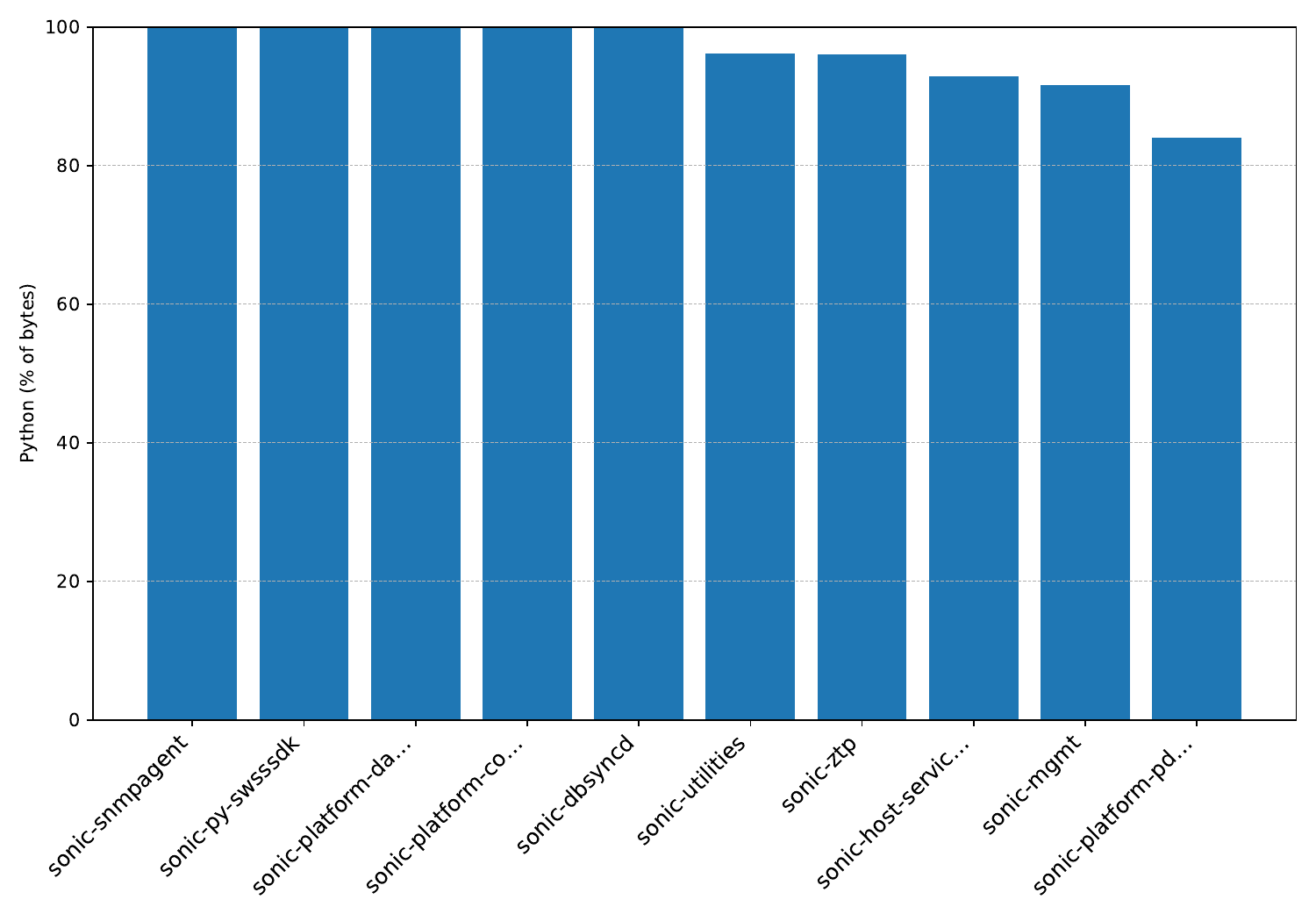}
    \caption{Prevalence of Python code in SONIC operating system repositories (percentage in bytes)}
    \label{fig:sonic-python-prevalence}
\end{figure}

Python applications were found to be part of the management and control pane in FastIron. These
were primarily installed on first boot following a firmware image update (or reflash). A persistent
flash partition was reserved for user data. This partition was proven to be persistent even across
factory reset events, with no facilities provided for removal and regeneration.

For example, firmware image \texttt{SPR09010jufi} contained 623 distinct PYC objects and 1499
standard Python source files. Among others, the TACACS subsystem relied on Python tooling, as well
as the web-based management UI, the inter-process communication protocol using Protocol Buffers
(a.k.a. protobuffs), and other critical components. The CPython interpreter, version 3.5.7, was integrated
into FastIron, alongside its standard libraries.

In addition, the removal or absence of PYC artifacts in the distributed firmware images is not
an indicator of intentional avoidance for security purposes, but two concomitant circumstances,
one case-specific, the other intrinsic to CPython:

\begin{itemize}
  \item FastIron integrity verification relies on compile-time "checksum" lists and file counts, only
  enforced in CC/FIPS operational modes. These checksums (in practice, cryptographic hashes, originally
  MD5, changed to SHA256 after the FLAPPYSWITCH disclosures) are calculated and integrated into the
  main core image inside the outer container image, during automated builds.
  \item PYC artifacts are not guaranteed to be reproducible \cite{pep552,python_design_faq}, and as such, it is very likely that any 
  subsequent mismatch between source files and their expected PYC counterparts would reflect in checksum
  mismatches, breaking boot during verification.
\end{itemize}

\begin{figure}[t]
  \centering
    \includegraphics[width=\linewidth]{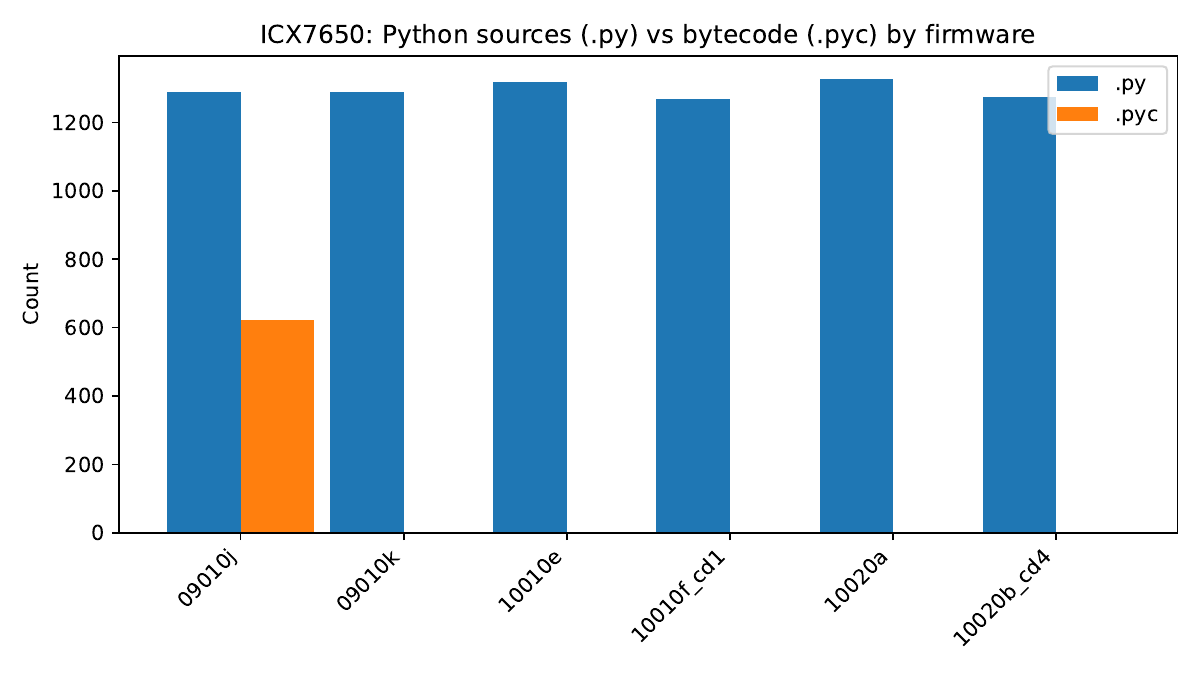}
    \caption{Prevalence of Python code in Ruckus ICX7650 FastIron firmware}
    \label{fig:icx7650-python-prevalence}
\end{figure}

Figure~\ref{fig:icx7650-python-prevalence} shows the prevalence of Python code
in FastIron firmware images for ICX7650 models. The aforementioned disappearance of PYC artifacts
is likely due to CI (firmware build system) configuration and the noted caveats.
The CPython interpreter will however honor any PYC artifacts in the flash partition
where the packages reside, thus enabling abuse of the preferential loading of compiled
bytecode files.

\subsubsection{Prevalence of Python code in the SONIC operating system}

The Software for Open Networking in the Cloud (SONiC) is a free and open source network operating
system based on Linux. It was originally developed by Microsoft and the Open Compute Project. It aims to
provide the entire software architecture required for a fully functional L3 (routing capable) device.
The project hosts its repositories at Github, facilitating review and code metrics.

The API offered by Github for auditing code language presence was used across the entire SONiC
organization to quantify Python code prevalence. Build-related repositories were explicitly excluded
through pattern-based matching. Figure~\ref{fig:sonic-python-prevalence} shows that core components
of the management and monitoring pane are Python-based.

Additionally, the firmware image artifacts from their automated CI build system were extracted
from the version targeting Broadcom-powered platforms. From the resulting \texttt{fs.squashfs}
SquashFS image, a total of 263 PYC artifacts and 5426 Python source files were counted, not including
any of the archives and packages additionally installed alongside the base system.

It is worth mentioning that among the packages provided, \textit{scapy} and \textit{paramiko} were
found to be installed in the firmware image. These provide network packet capture and manipulation,
and programmatic SSH session capabilities, respectively. Both of these would be readily usable and
versatile for adversarial lateral movement in the network, as well as traffic capture, for example
to collect NTLM hashes (or similar ticketing protocol tokens), effectively enabling "living off
the land" so-called tools and techniques within core network infrastructure.

\subsection{Persistent vectors via Python bytecode infection}
\label{sec:python-bytecode-infection}
\label{s:persistence}

Past research on Python bytecode infection has mostly revolved around targeting
the marshalled Python code itself, that is, the PyCodeObject stream \cite{BytecodeBytecode0171dev9+g13032bbdc}. Literature
in academic resources in this domain is scarce, but industry and hacking scene
researchers have explored this vector in the past. Typical procedures---in the
less sophisticated form---involve concatenation of a new payload alongside the
original compiled source, with manipulation of the \path{LOAD_CONST} data, injected
into the \path{co_consts} of the final bytecode stream product. More sophisticated
manipulation techniques revolve around fine-grained manipulation of bytecode
opcodes and data, leveraging existent functionality in an ad hoc fashion to
achieve opportunistic loading of adversary-controlled external code components.

Regardless, in order to achieve persistence, more so in post-exploitation scenarios,
the loading strategy of the interpreter, alongside the opportunities for mutable---and, crucially, persistent---storage
manipulation, become the necessary building blocks.

Because the interpreter will attempt to leverage environment-influenced loading paths,
and the deployment of the CPython environment is on its own highly specific to
each use case (as the FastIron vulnerabilities exemplified), the interpreter
cannot offer integrity assurances. Through careful placement of PYC artifacts
and abuse of the weak invalidation schemes, original Python sources can be
side-loaded after the adversary-controlled PYC(s) have already executed in the
CPython bytecode virtual machine, retaining original functionality while the
adversary-controlled code is preferentially loaded.

The FLAPPYSWITCH vulnerabilities were demonstrated in the form of a multi-stage
loader payload that leverages unused NAND flash partitions for storing adversary
payloads, as well as persistent data partitions (as used for configuration and certificates).
This is precisely the technique that can be generalized, especially in the context
of critical infrastructure equipment, presenting a significant, if not often insurmountable,
forensic challenge.

With the increasing prevalence of Python artifacts in the control and management pane
of industrial, enterprise networking and communications devices, and the increasing difficulty in
exploitation of traditional memory corruption vulnerabilities, these techniques will
only become more refined and common in the field, as both execution \cite{9474291,10179370} and evasion \cite{valerosStudyMacheteCyber2019} substrate,
with substantial evidence of Python being normalized as an operational substrate in
industrial attack tooling \cite{di2018triton}.

Last but definitely not least, because of the single-bit invalidation ``disable'' switch,
a far more inconspicuous form of persistence is achieved through forcing a vulnerable
compiled version of a file to be loaded in place of one that fixes or mitigates a
vulnerability. For example, should a Python package contain a vulnberability in one of its
source files, and a subsequent patch be released by its maintainers, an adversary could leave
a PYC in place, with the exploitable condition itself still present as means to re-attain
access via the same vulnerability, despite apparent mitigation in source.

\section{Architecture considerations of bytecode loading and trust chain impact}
\label{s:architecture}

Presently, Trust Chain decisions and integrity models in operating systems revolve primarily
around limiting native code execution \cite{5207638} and loading, as measurable components that can be signed
or {\em pinned} based off path or filesystem properties---such as inodes--, where the semantics
of reading and executing an object---or artifact---are clearly delimited and enforceable by policy
(eg. SELinux\cite{selinux}, early TPE modules in the Linux kernel, etc)~\footnote{Furthermore, current computing
architectures have also implemented protection---or permission- semantics for memory management,
where read does not imply execute anymore. This does not hold for loaded scripts, bytecode or
interpreter programs where read access implies execution}. These policies control execution of
native code but break down at the point of the interpreter loading arbitrary files (or network
streams)~\footnote{For instance, PHP remote inclusion vulnerabilities have been a staple of web
application insecurity, where the interpreter internally loads remote code, manifesting originally---from the side of the operating system---as a network socket operation and descriptor I/O, otherwise
invisible to OS---and especially kernel-level---security decision making.},
that can then execute as code or be loaded as such. In order to interoperate with TPE, instead of
breaking its guarantees, interpreters must be explicitly modeled within the trust architecture
of the system and provide means for fine-grained control of their code loading behaviors.

Bytecode loading by the CPython interpreter is performed without any Trusted Path Execution (TPE)
considerations. The challenges of securing dynamic language runtimes have been well documented
in academia and industry literature \cite{10.1145/2500727.2500747}, with MITRE ATT\&CK T1059
(Command and Scripting Interpreter Sub-techniques) \cite{CommandScriptingInterpreter} as a disctintive
class of techniques involving such runtimes becoming prevalent in adversarial scenarios.
To date, there is no explicit published research in the domain of applying Trusted Path Execution (TPE)
principles to dynamic language runtimes, such as script interpreters.

As a result, CPython does not impose security restrictions based
on privilege or integrity when loading \texttt{.pyc} files. Any Python module
with a writable bytecode cache can therefore serve as a vector for persistent
malicious code, as explained in Section~\ref{sec:python-bytecode-infection}.

Since no native Python runtime provisions exist to establish a verifiable trust chain, 
ad hoc coupling of existent OS-dependent technologies, such as \texttt{dm-verity},
can be used for filesystem or path-bound object integrity monitoring.
Such file-level monitoring and verification, however, require non-trivial
considerations as every Python artifact is a potential vector for untrusted code
loading.

Moreover, systems like \texttt{dm-verity} have been impacted by completely
unrelated vulnerabilities in other OS subsystems that were not scoped in the
threat model of its developers, such as the \textit{LoadPin} bypass vulnerability
\cite{LinuxLoadPinBypass}.

Successive changes to the \texttt{.pyc} header format attempted to improve
correctness but did not introduce true enforceable security guarantees. Timestamp
validation relies on mutable filesystem metadata, while hash-based validation
(PEP~552\cite{pep552}) remains optional and locally scoped. We describe an additional vector
requiring minimal header manipulation in Section~\ref{sec:unchecked-hash-bypass}.

\section{Future work: Designing enforceable TPE-like policies for Python}
\label{s:future}

To implement enforceable TPE-like policies for Python code loading, careful evaluation of
the CPython import system and runtime behaviors is required.

Based on analysis of the CPython interpreter source as of January~2025, the following
roadmap outlines a layered implementation strategy for prototyping a trusted Python
loader. The objective is to introduce security decision-making capabilities analogous
to Trusted Path Execution (TPE) at points in the import control flow where sufficient
context is available \emph{prior} to code loading or execution, while preserving
deterministic semantics and minimizing performance impact.

Successful implementation of a prototype hinges on establishing a precise threat model and define what constitutes trusted code,
trusted storage, and trusted provenance. This includes specifying fail-closed semantics,
explicitly disallowing ambiguous behavior such as silent fallback from invalid bytecode
to source compilation, and formalizing system-wide invariants:

\begin{itemize}
  \item no execution of unauthenticated code objects,
  \item all policy decisions occur prior to unmarshalling or execution,
  \item integrity ambiguity results in rejection rather than recovery, and
  \item enforcement relies on stable identifiers (cryptographic or filesystem anchors)
  rather than mutable path strings. These invariants serve as correctness criteria
  throughout the implementation.
\end{itemize}

In order to introduce policy decision hooks at context-rich import boundaries
\footnote{
  The challenge of adequate coverage for hooks in security frameworks has long been a
  subject of study, including the seminal work by \citeauthor{smalley2001implementing,wright2002linux,spengler2005increasing} \cite{smalley2001implementing,wright2002linux,spengler2005increasing}.
  Placement is critical to ensure that all relevant events are captured without excessive overhead,
  but also avoiding the introduction of exploitable gaps or unexpected interactions---or opportunities
  for such---in the process.
}
, the following placement candidates are proposed:

\begin{itemize}
  \item \emph{Finder-stage provenance hook:} Instrument
  \path{PathFinder.find_spec} in \path{importlib._bootstrap_external} to capture
  the resolving \texttt{sys.path} entry, resolved origin, loader class, and package
  context. This stage provides stable provenance information that is otherwise lost
  once resolution completes.
  \item \emph{Loader-stage pre-execution hook:} Instrument
  \path{SourceFileLoader.get_code} and
  \path{SourcelessFileLoader.get_code} to invoke policy checks immediately before
  reading source files or unmarshalling bytecode. This represents the primary trusted
  path boundary, where rejection can occur without partial execution.
\end{itemize}

These hooks naturally reside in the importlib bootstrap code, with optional C-level
enforcement in \texttt{import.c} to harden against runtime bypass.

\subsection{Beyond hooks: hardening loader and execution semantics}

\subsubsection{Selecting a PYC integrity representation strategy}
Determine how cryptographic integrity metadata is associated with bytecode artifacts:
\begin{itemize}
  \item \emph{Out-of-band sidecar signatures:} Preserve the existing \texttt{.pyc}
  format and store detached signatures or MACs alongside cache entries. This minimizes
  compatibility risk and isolates cryptographic handling from marshal parsing.
  \item \emph{In-band header or trailer extensions:} Extend the \texttt{.pyc} container
  with new integrity fields and flags, requiring explicit parsing to ensure only the
  code object payload is unmarshalled when integrity assurances hold true.
\end{itemize}

In either case, the authenticated region must be clearly defined and downgrade or
stripping attacks must be prevented by policy.

\subsubsection{Enforcing cryptographic verification prior to unmarshalling}
Integrate integrity verification immediately after reading \texttt{.pyc} bytes and
before any header interpretation or \path{marshal.loads} invocation. This step must
be mandatory for:
\begin{itemize}
  \item \emph{Sourceless imports}, where no source file is available for cross-checking.
  \item \emph{Source-backed imports}, whenever cached bytecode is selected for loading.
\end{itemize}
Failure to verify integrity must result in immediate rejection unless explicitly
permitted by policy.

\subsubsection{Hardening PYC--source mismatch handling with fail-closed semantics}
While CPython already detects timestamp- and hash-based mismatches \cite{vuLastPyMileIdentifyingDiscrepancy2021}, the prototype
should surface these events to the trusted agent and enforce consistent outcomes.
In hostile environments, mismatch should be treated as an integrity violation rather
than a trigger for recompilation. Optional policy-controlled fallback modes may be
supported for development or controlled build environments, but must be explicit and
auditable. The current check-hash default mode should issue warnings rather than silently
accepting bytecode when PYC-source equality cannot be established (as described in
Section~\ref{sec:unchecked-hash-bypass}).

\subsubsection{Integrate a trusted decision agent operating on stable anchors}
Design a narrow interface through which the import system communicates with a trusted
agent, supplying module identity, resolved origin, loader and finder metadata,
\path{sys.path} provenance, filesystem identifiers (device, inode, permissions),
and integrity verification results. Policies should primarily rely on cryptographic
identities or immutable filesystem properties rather than path names. The agent may
be embedded within the interpreter for stronger enforcement or externalized for
flexibility, but the decision point must remain in-process and fail closed.

\subsubsection{Add PKI and key management support for supply chain coverage}
To address supply chain threat models \cite{vuLastPyMileIdentifyingDiscrepancy2021,cesarano2025goleash}, cryptographic verification must be paired with
trust distribution mechanisms, including publisher trust anchors, key rotation, and
revocation. Transport security alone is insufficient to mitigate compromised
distribution infrastructure or malicious publishers. Missing or untrusted keys must
be treated as policy-relevant conditions rather than implicit acceptance.

\subsubsection{Harden enforcement against bypass and ensure auditability}
Introduce interpreter-level controls to ensure that policy hooks cannot be skipped or
disabled through runtime modification. Storing policy callbacks in interpreter state
and invoking them from fixed choke points mirrors existing audit-hook mechanisms.
Structured logging of import decisions and integrity outcomes should be included to
support forensic analysis and incident response.

\subsubsection{Validate correctness, coverage, and performance impact}
Develop a comprehensive test matrix covering signed and unsigned bytecode, tampered
caches, mismatch scenarios, symlink and path-substitution attacks, virtual
environments, namespace packages, and alternative loaders. Validate that invariants
hold system-wide, denial behavior is deterministic, and performance overhead remains
within acceptable bounds for import-heavy workloads.

\section{Related work}

There exists considerable prior research in understanding Python interpreter internals, bytecode
generation and invalidation, malware persistence mechanisms, and embedded system
firmware analysis. However, these directions largely did not intersect and thus overlooked
Python's loading and bytecode invalidation as an attack surface.

The paper \cite{duanMeasuringSupplyChain2021} considers supply chain integrity risks of
package managers for interpreted languages, and proposes a framework to evaluate their
security features. We address the risk and integrity problem for the interpreters themselves.

\subsection{Python bytecode and import system semantics}

The structure, generation, and loading of Python bytecode have been documented
primarily for reasons of correctness, performance, and compatibility.
PEP~3147~\cite{pep3147} and PEP~3149~\cite{pep3149} formalized repository layouts
and ABI tagging to support multi-version deployments, while PEP~552~\cite{pep552}
introduced deterministic and hash-based invalidation to address reproducibility
and build determinism. Canonical documentation of the import system and bytecode
execution model appears in CPython source code and language references
\cite{cpython_importlib_bootstrap,cpython_import,python_language_reference}.

Importantly, these sources consistently frame PYC files as cache
artifacts whose validation mechanisms are intended to prevent stale execution,
not to enforce integrity or trust boundaries. The unchecked-hash mode introduced
by PEP~552 is explicitly documented as an optimization and deployment feature,
not as a security control.

\subsection{Supply chain risks}
\label{sec:supply-chain-risks}

Prior work has pointed out supply chain risks of divergent views of source and
bytecode by Python's runtimes---and other interpreters or dynamic language runtimes \cite{duanMeasuringSupplyChain2021, rayarao2025shai}. For instance, 
the popular PyPi distribution system exhibits these risks.~\cite{vuLastPyMileIdentifyingDiscrepancy2021,WhenByteCode2023a,gao2025malguardrealtimeaccurateactionable}

We note that the risks extend beyond avoiding source code scanning, to malicious persistence \cite{WhenByteCode2023a}.

\subsection{Pickling and user-provided serialized objects}

Attacks via compiled cached bytecode have a dual: attacks via maliciously crafted
serialized (a.k.a. pickled) data. It is important to distinguish between the two.  In the
bytecode case, behaviors of the Python's loader are targeted and the loader is deputized
by the attacker. In the pickled data case, Python's deserialization functionality provided
by the \texttt{pickle} module is targeted.

Prior work involving Python interpreter safety has primarily focused on the risks associated with deserializing 
untrusted pickled data, leveraged against the \texttt{pickle} module and libraries
depending on its serialization features. However, these
concerns are distinct from bytecode integrity issues, as they involve a different serialization
mechanism and are typically mitigated through careful use of the \texttt{pickle} module's security
features. A natural extension of this domain is safety of user-provided models, as is the case with
Hugging Face Model Hub and similar machine learning model exchanges \cite{kellasPickleBallSecureDeserialization2025}.

\subsection{Fickling: defensive and offensive approaches for PyTorch and related tooling}

Perhaps the most practical and comprehensive research effort in this domain is "Fickling" \cite{sultanikNeverDillMoment2021},
performed outside of academia and presented at Defcon 2021. \textit{Fickling} is a tool designed to analyze and manipulate
Python bytecode, offering both defensive and offensive capabilities for PyTorch and related tooling \cite{TrailofbitsFicklingPython}.

For defensive purposes, it provides basic classification capabilities to detect or score possible
malicious payloads inside a given object. This is likely insufficient against language-specific
techniques, not dissimilar to ROP (Return Oriented Programming), as demonstrated in
\cite{liuArtHideSeek2025} and \cite{huangPainPickleBypassing2022}.

\subsection{Post-compromise persistence}

To our knowledge, this is the first work to systematically analyze CPython’s
bytecode invalidation modes as an attack surface for post-compromise persistence.
Rather than introducing new bytecode manipulation techniques, we demonstrate
that existing, documented interpreter behavior is sufficient to make modified
PYC files authoritative under realistic threat models. This reframing
connects previously isolated domains---Python internals, malware persistence, and
embedded system deployment---into a unified security analysis.

\section{Conclusion}

Owing to increasing penetration of Python into computing infrastructure devices, treating Python's
ecosystem as a part of these devices' trust chain is necessary. Python loaders must now be
analyzed and enhanced for trust chain policy enforcement just as other loaders that precede them
in the traditional chains of trust. Trusted Path Execution (TPE) policies must become enforceable
in the Python ecosystem.

Moreover, a key challenge of Python is that its different components have multiple loading behaviors
and interactions---modules and libraries themselves depending on or introducing
second and third-order interactions\footnote{Python modules and libraries themselves depending on or introducing
additional runtime code loading vectors, from marshalling or serialization to FFI and native
code dependencies.}---dramatically increasing the complexity of successfully
locking down the interpreter with a TPE-like policy against malicious runtime code loading and injection.
In effect, this requires defining the concept of ``Trusted Loader Execution'' (TLE), a generalization
of the TPE.

\bibliographystyle{ACM-Reference-Format}
\bibliography{cpython-pyc-integrity-checking}

\end{document}